\documentclass[prd,reprint,nofootinbib,superscriptaddress]{revtex4-2}
\usepackage{bm}
\usepackage{ulem}
\usepackage{comment}
\usepackage{amsmath}
\usepackage{amssymb}
\usepackage{booktabs}
\usepackage{graphicx}
\usepackage{float}
\usepackage{comment}
\usepackage{hyperref}
\usepackage{CJK}
\usepackage{color}
\hypersetup{colorlinks=true, linkcolor=red, citecolor=blue}

\begin{document}
\begin{CJK*}{UTF8}{gbsn}
\title{Testing alternative theories of gravity with space-based gravitational wave detectors}
\author{Qing Gao (郜青)}
\email{gaoqing1024@swu.edu.cn}
\affiliation{School of Physical Science and Technology, Southwest University, Chongqing 400715, China}
\author{Yujie You (尤雨婕)}
\affiliation{School of Physical Science and Technology, Southwest University, Chongqing 400715, China}
\author{Yungui Gong (龚云贵)}
\email{yggong@hust.edu.cn}
\affiliation{School of Physics, Huazhong University of Science and Technology, Wuhan, Hubei
430074, China}
\author{Chao Zhang (张超)}
\email{zhangchao666@sjtu.edu.cn}
\affiliation{School of Physics and Astronomy, Shanghai Jiao Tong University, Shanghai 200240, China}
\author{Chunyu Zhang (张春雨)}
\email{Corresponding author. chunyuzhang@itp.ac.cn}
\affiliation{CAS Key Laboratory of Theoretical Physics, Institute of Theoretical Physics, Chinese Academy of Sciences, Beijing 100190, China}

\begin{abstract}
We use gravitational waves (GWs) from binary black holes (BBHs) and neutron stars inspiraling into intermediate-mass black holes to evaluate how accurately the future space-based GW detectors such as LISA, Taiji and TianQin and their combined networks can determine source parameters and constrain alternative theories of gravity.
We find that, compared with single detector, the detector network can greatly improve the estimation errors of source parameters, especially the sky localization,
but the improvement of the constraint on the graviton mass $m_g$ and the Brans-Dicke coupling constant $\omega_{BD}$ is small.
We also consider possible scalar modes existed in alternative theories of gravity and
we find the inclusion of the scalar mode has little effect on the constraints on source
parameters, $m_g$, and $\omega_{BD}$ and the parametrized amplitude $A_B$ of scalar modes are small.
For the constraint on the graviton mass,
we consider both the effects in the GW phase and the transfer function due to the mass of graviton.
With the network of LISA, Taiji and TianQin,
we get the lower bound on the graviton Compton wavelength $\lambda_g\gtrsim 1.24 \times 10^{20}$ m
for BBHs with masses $(10^6+10^7)M_\odot$, and $A_B< 5.7\times 10^{-4}$ for BBHs with masses $(1+2)\times 10^5M_\odot$;
$\omega_{BD}>6.11\times10^{6}$
for neutron star-black hole binary with masses  $(1.4+400)M_{\odot}$.
\end{abstract}


\maketitle
\end{CJK*}

\section{Introduction}

The discovery of the accelerating expansion of the Universe by the observations of type Ia supernovae \cite{SupernovaSearchTeam:1998fmf,SupernovaCosmologyProject:1998vns} in 1998 imposes a great challenge on Einstein's general theory of relativity (GR).
The late-time cosmic acceleration together with the problems of dark matter, quantum gravity, and singularity of spacetime might imply that GR needs to be modified.
The simplest extension to GR is to add a scalar field such as Brans-Dicke (BD) theory \cite{Brans:1961sx}.
The BD field $\phi$ not only plays the role of Newton's gravitational constant but also mediates gravitational interaction.
The most general scalar-tensor theory that leads to at most second-order field equations is Horndeski theory \cite{Horndeski:1974wa} and its generalizations \cite{Deffayet:2009mn}.
Another modification of GR is to introduce higher-order terms of the Riemann tensor in the action such as
the general nonlinear $f(R)$ gravity \cite{Buchdahl:1970ynr} and Lovelock gravity \cite{Lovelock:1971yv}.
If graviton has a mass $m$, then gravity is screened at a scale larger than $m^{-1}$, henceforth, massive gravity theory may explain the cosmic acceleration.
Of course, the mass of graviton should be small so that gravity can be approximated as a long-range force.
However, the linearized Fierz-Pauli (FP)  massive gravity \cite{Fierz:1939ix} suffers the problem of the van Dam, Veltman and Zakharov (vDVZ) discontinuity \cite{vanDam:1970vg,Zakharov:1970cc}.
Recently, the vDVZ problem was overcome in the generalized linear massive gravity \cite{Gambuti:2021meo} without using the Vainshtein mechanism \cite{Vainshtein:1972sx}. 
In Dvali, Gabadadze and Porrati (DGP) gravity \cite{Dvali:2000hr},
GR is modified at the cosmological scale and there is a continuous tower of massive gravitons.
In the de Rham, Gabadadze, and Tolley (dRGT) massive gravity \cite{deRham:2010kj},
higher-order nonlinear terms were introduced and the theory is free from Bouldware-Deser ghost \cite{Boulware:1972zf,Hassan:2011hr}.
Other alternative theories of gravity include
the tensor-vector-scalar theory, or T$e$V$e$S theory \cite{Bekenstein:2004ne} and the generalized T$e$V$e$S theory \cite{Seifert:2007fr},  Einstein-\ae ther theory \cite{Jacobson:2000xp,Jacobson:2004ts}, dynamical Chern-Simons gravity \cite{Jackiw:2003pm}, Ho\v{r}ava gravity \cite{Horava:2009uw}
and bimetric theory of gravity \cite{Rosen:1940zz,Rosen:1973zz}, etc.
For reviews of alternative theories of gravity, see Refs. \cite{Nojiri:2010wj,Hinterbichler:2011tt,Clifton:2011jh,deRham:2014zqa,Nojiri:2017ncd}.

The constraint on the mass of graviton from the new solution of the ephemeris INPOP19a is $m_g\leq 3.16\times 5.63\times 10^{-59}$ kg at the 90\% confidence level \cite{Bernus:2020szc}.
Using the measurements on the secular change of the orbital period from PSR B1913+16 \cite{Weisberg:2016jye} and PSR J1738+0333 \cite{Freire:2012mg},
the constraints on the  mass of graviton in FP gravity \cite{Fierz:1939ix,Gambuti:2021meo} and DGP gravity \cite{Dvali:2000hr} were obtained in Ref. \cite{Poddar:2021yjd}.
In BD theory \cite{Brans:1961sx,eardley1975},
the extra dipolar radiation further decreases the orbital period of a binary system \cite{eardley1975,Will:1977wq},
so the measurement on the secular change in the orbital period of a compact binary can constrain the BD coupling constant \cite{eardley1975,Will:1977wq,Will:1989sk,Damour:1992we,Damour:1998jk,Alsing:2011er,Antoniadis:2013pzd,Zhang:2019hos,Seymour:2019tir}.
By using the observational results from the pulsar-white dwarf binary PSR J1738+0333,
it was obtained that  $\omega_{\text{BD}}> 25000$ \cite{Freire:2012mg}.
The  Cassini measurement \cite{Bertotti:2003rm} on the Shapiro time delay in the solar system
gave the most stringent constraint on BD coupling constant $\omega_{\text{BD}} > 40000$ \cite{Bertotti:2003rm,Will:2014kxa}.

The detection of gravitational waves (GWs) from binary black holes (BBHs) by
the Laser Interferometer Gravitational-Wave Observatory (LIGO) Scientific Collaboration and the Virgo Collaboration
opened a new window to test GR and probe the nature of gravity in the strong-field and nonlinear regions \cite{TheLIGOScientific:2016src,Abbott:2018lct,LIGOScientific:2019fpa,Abbott:2020jks}.
So far, there are tens of detected GW events, and we are in a new era of gravitational Universe \cite{Abbott:2016blz,LIGOScientific:2018mvr,LIGOScientific:2020ibl,LIGOScientific:2021usb,LIGOScientific:2021djp}.
The propagation speed of GWs in GR is the same as the speed of light and there are two
transverse tensor polarizations.
In alternative theories of gravity, the propagation speed of GWs may differ from the speed of light, and GWs may have up to six polarizations \cite{Eardley:1974nw,Liang:2017ahj,Hou:2017bqj,Gong:2017bru,Gong:2017kim,Gong:2018cgj,Gong:2018ybk,Gong:2018vbo,Hou:2018djz},
so the measurement of polarization states can distinguish alternative theories of gravity from GR \cite{Nishizawa:2009bf,Hayama:2012au,Isi:2015cva,Isi:2017equ,Callister:2017ocg,DiPalma:2017qlq,LIGOScientific:2018czr,Takeda:2018uai,Takeda:2019gwk,Liu:2020mab,Zhang:2021fha}.
In massive gravity, the propagation speed of GWs with higher frequency is larger,
so GWs emitted at earlier time with lower frequencies may reach the detector later than GWs emitted at later time with higher frequencies and the observed phase of GWs will be distorted;
henceforth, the observed dephasing in GW waveforms can place a bound on the mass of graviton.
Using the TaylorF2 waveform model and the low-spin prior,
the Bayesian analysis of the first binary neutron star (NS) event GW170817 gave $m_g\le 1.305\times 10^{-57}$ kg \cite{Shoom:2022cmo},
which is weaker than the 90$\%$ credible bound $m_g\leq 2.26\times 10^{-59}$ kg
\cite{LIGOScientific:2021sio} placed by the LIGO, Virgo and KAGRA collaborations with the IMRPhenomXP waveform model \cite{Garcia-Quiros:2020qpx} by using 43 selected GWTC-3 events including 12 BBHs observed during the second half of the third observing run.
Considering the GW events with a false alarm rate (FAR) less than $10^{-3}$ yr$^{-1}$ and signal-to-noise (SNR) $\ge 6$ in the inspiral stage from all 15 GW events observed during the second half of the third observing run
and using parametric deviations to the GW inspsiral phase in the frequency-domain model SEOB-NRv4\_ROM \cite{Bohe:2016gbl,Cotesta:2020qhw},
it was found that all post-Newtonian deformation coefficients are consistent with the predictions of GR and the bound on $\delta\hat\varphi_{-2}$ was improved by a factor of $\sim 2$ upon the GWTC-2 bound due to the long duration of the NS/black hole (BH) event GW200115\_042309 \cite{LIGOScientific:2021sio}.
Based on an effective antenna pattern function that is constructed from a subset of polarization modes
and performing the null projection with respect to the subspace spanned by the component of the beam pattern vectors parallel to the basis mode(s),
no evidence of departing from the pure tensorial hypothesis as predicted by GR was found from all the events from the first three observing runs with a FAR less than $10^{-3}$ yr$^{-1}$ threshold \cite{LIGOScientific:2021sio}.
Because of the no-hair theorem,
BHs do not carry scalar charges
and the orbital evolution and gravitational radiation from BBHs
are identical in GR and BD theory,
so we cannot use BBHs to distinguish BD theory from GR via dipole radiation \cite{Will:1989sk,Yunes:2011aa,Mirshekari:2013vb}.
Considering a NS with mass $0.7M_\odot$ inspiraling into a BH with mass $3M_\odot$ for the SNR $\rho=10$,
LIGO/Virgo detectors could give a bound $\omega_{\text{BD}}\gtrsim 2000$ with the Fisher information matrix (FIM) method \cite{Will:1994fb}.
However, the observation of the GW event GW190426\_152155 of a possible $1.5M_\odot$ NS/$5.7M_\odot$ BH binary by LIGO and Virgo placed the bound $\omega_{\text{BD}}\gtrsim 10$
using the Bayesian inference method \cite{Niu:2021nic}.

The ground-based GW detectors can detect GWs from the coalescence of stellar-mass compact binaries in the frequency range $10-10^3$ Hz only,
but the future space-based GW detectors such as LISA \cite{Danzmann:1997hm,LISA:2017pwj},
TianQin \cite{TianQin:2015yph} and Taiji \cite{Hu:2017mde,Gong:2021gvw} can detect GWs in the millihertz frequency band with long-duration signals from larger distance.
It is expected that GW detection from inspiral to merge and ringdown with LISA, Taiji or TianQin will place
more stringent constraints on alternative theories of gravity.
For one-year observations of BBHs with masses $(10^7+10^7)M_\odot$ at the luminosity distance $D_L=3$ Gpc prior to the innermost stable circular orbit (ISCO),
LISA could place the bound on the graviton Compton wavelength $\lambda_g=h/(m_g c)$ as $\lambda_g\geq 6.9\times 10^{19}$ m  by using the pattern-averaged detector noise curve and GW waveform up to 1.5 post-Newtonian (PN) order with the FIM method \cite{Will:1997bb};
employing the LISA sensitivity curve generator, the lower bound was revised as $\lambda_g>4.8\times10^{19}$ m \cite{Will:2004xi};
using pattern-averaged templates at 2PN order and considering the spin-orbit coupling,
LISA could give the lower bound $\lambda_g>2.2 \times10^{19}$ m \cite{Berti:2004bd}.
Instead of taking the pattern average,
by Monte Carlo simulations of $10^4$ BBHs randomly distributed and oriented in the sky with mass $(10^6+10^6)M_{\odot}$,
the average lower bound $\lambda_g>1.33\times10^{19}$ m with LISA was given in \cite{Berti:2004bd}.
For one-year observations of precessing eccentric BBHs  at $D_L=3$ Gpc prior to the ISCO,
applying the restricted 2PN GW waveforms including the effects of spin-orbit and spin-spin couplings, spin precession and eccentricity,
LISA could give the constraint $\lambda_g>1.2\times10^{19}$ m
with the pattern-average analysis on $(10^7+10^7)M_{\odot}$ BBHs,
while the Monte Carlo simulations of $10^4$ BBHs randomly distributed and oriented in the sky with masses $(10^7+10^6)M_{\odot}$ by LISA gave the average bound $\lambda_g>3.1\times10^{19}$ m \cite{Yagi:2009zm}.
In these studies, the effect of the propagation speed of the massive graviton in the transfer function was not considered.
In Ref. \cite{Gao:2022hho}, the author took this effect into account and proposed an approximate analytical expression for the averaged transfer function, and also considered the accuracy of parameter estimation with different space-based GW detectors.
It was found that the effect of the transfer function due to massive graviton on parameter estimation for small graviton mass is negligible,
and for one-year observations of BBHs with masses $(10^6+10^6)M_{\odot}$ prior to the ISCO using 3.5PN GW waveform and the FIM method,
Taiji could give better constraint than LISA and TianQin and the result is $\lambda_g>1.91\times 10^{19}$  m \cite{Gao:2022hho}.
For one-year observations of a $1.4M_\odot$ NS inspiraling into a $10^3 M_\odot$ BH prior to the ISCO with a SNR of 10,
using the pattern-averaged GW waveforms at 1.5PN order and the FIM method,
LISA could give the constraint  $\omega_{\text{BD}}\gtrsim 244549$  \cite{Scharre:2001hn},
or $\omega_{\text{BD}}\gtrsim 203772$ with the detector noise curve from  the LISA sensitivity curve generator \cite{Will:2004xi};
with the addition of the spin-orbit coupling to the pattern-averaged 2PN GW templates,
the bound with LISA became $\omega_{\text{BD}}\gtrsim 21257$ \cite{Berti:2004bd},
while $\omega_{\text{BD}}\gtrsim 10799$ by taking the average of $10^4$ randomly distributed binaries  with a SNR of $\sqrt{200}$ \cite{Berti:2004bd};
including the spin-orbit and spin-spin couplings to the pattern-averaged restricted 2PN GW waveforms for precessing eccentric orbits,
LISA could give the bound $\omega_{\text{BD}}\gtrsim 1881$ \cite{Yagi:2009zm},
and $\omega_{\text{BD}}\gtrsim 3523$
using the average of $10^4$ randomly distributed binaries with a SNR of $\sqrt{200}$ \cite{Yagi:2009zm}.
By numerically calculating the eccentric orbits of a $1.4M_\odot$ NS inspiraling into a massive BH with the methods of osculating orbits and self-force,
LISA could give the constraint $\omega_{\text{BD}}>10^6$ using the mismatch of GW waveforms between GR and BD theory with two-year observations \cite{Jiang:2021htl}.

It is well known that extra polarization states exist in both massive gravity and BD theory,
but the effects of extra polarization states on parameter estimation were not discussed in the above work.
In the present work, we consider the impact of scalar modes on parameter estimation by parametrizing the scalar mode in massive gravity $h_B^{\ell m}(f)$ as $A_{B} h_{\ell -m}(f) Y_{\ell m}(\iota,0)$ \cite{Chatziioannou:2012rf,Isi:2015cva,Takeda:2018uai,Takeda:2019gwk,Pang:2020pfz,Wong:2021cmp,Zhang:2021fha}.
Note that in the above studies only the observations with a single detector like LISA was used to constrain massive gravity and BD theory,
so we study how accurately LISA, Taiji and TianQin and their combined networks can determine source parameters and constrain alternative theories of gravity
by Monte Carlo simulation of 1000 randomly distributed and oriented binaries in the sky,
because the detector network can greatly improve the sky localization compared with a single detector \cite{Gong:2021gvw,Ruan:2020smc,Zhang:2020hyx,Zhang:2020drf,Zhang:2021kkh,Zhang:2021wwd}.
Although it was found that the effect of the mass of graviton on the averaged transfer function is negligible \cite{Gao:2022hho},
we do not neglect the effect of the graviton mass in the antenna response.
In this paper, we employ the IMRPHENOMXHM waveform model \cite{Garcia-Quiros:2020qpx} which includes higher harmonic modes to make the parameter estimation.
The paper is organized as follows. In Sec. \ref{sec2},
we discuss the antenna response and the parameter estimation with the FIM method.
In Sec. \ref{sec3}, we discuss the estimate errors of source parameters and
the constraints on $\lambda_g$ and $\omega_{BD}$ with LISA, Taiji, and TianQin and their combinations.
The conclusion is drawn in Sec. \ref{sec4}.
In this paper, we set $G=c=h=1$.

\section{Fisher information matrix method}
\label{sec2}
In terms of the polarization tensor $e^A_{ij}$,
GWs can be expressed as
\begin{equation}
\label{hijt}
h_{ij}(t)=\sum_{A} e^A_{ij} h_A(t),
\end{equation}
where $A=+,\times,x,y,b,l$ stands for the plus, cross, vector $x$,
vector $y$, breathing, and longitudinal polarizations, respectively.
For GWs propagating in the  direction $\hat{\Omega}$,
the signal registered in the GW detector is
\begin{equation}
\label{gwst}
s[h(t)]=\sum_A F^A h_A(t),
\end{equation}
where the angular response function $F^A$ for the polarization $A$ is
\begin{equation}
\label{faeq1}
F^A=\sum_{i,j} D^{ij} e^A_{ij},
\end{equation}
the detector tensor $D^{ij}$ for an equal arm space-based interferometric detector with a single round-trip light travel is
\begin{equation}
D^{ij}=\frac{1}{2}\left[\hat{u}^i \hat{u}^j T(f,\hat{u}\cdot\hat{\Omega})-\hat{v}^i \hat{v}^j  T(f,\hat{v}\cdot\hat{\Omega})\right]e^{i\Phi_D},
\end{equation}
$\hat{u}$ and $\hat{v}$ are the unit vectors along the two arms of the detector,
\begin{equation}
    \Phi_D(t)=2\pi fR_e\sin(\theta_s)\cos(\omega_e t-\phi_s+\varphi_i)/c,
\end{equation}
is the Doppler shift,
$R_e=1\ {\rm A.U.}$ is the orbital radius, $(\theta_s,\phi_s)$ is the source location,
$\omega_e=2\pi/T_e$ is the orbital frequency of the Earth, $T_e=1$ year is the period, and $\varphi_i$ is the ecliptic longitude of the detector at $t=0$.
In massive gravity theories,
the propagation speed of GWs $v_{gw}(f)=\sqrt{1-(m_g c^2)^2/(2\pi \hbar f)^2}$ is different from the speed of light $c$ and depends on the frequency of GWs.
Taking this effect into account,
the transfer function $T(f,\hat{n}\cdot\hat{\Omega})$ for a single round trip in the arm is \cite{Tinto:2010hz,Blaut:2015qaa}
\begin{equation}
\label{transferfunction}
\begin{split}
T(f,\hat{n}\cdot\hat{\Omega})=\frac{1}{2}& \left\{ \text{sinc}\left[\frac{f}{2f^*}(1-\hat{n}\cdot\hat{\Omega}/(v_{gw}/c))\right]\times\right.\\
&\exp\left[-i\frac{f}{2f^*}(3+\hat{n}\cdot\hat{\Omega}/(v_{gw}/c))\right] \\
& +\text{sinc}\left[\frac{f}{2f^*}(1+\hat{n}\cdot\hat{\Omega}/(v_{gw}/c))\right]\times\\
&\left.\exp\left[-i\frac{f}{2f^*}(1+\hat{n}\cdot\hat{\Omega}/(v_{gw}/c))\right]\right\},
\end{split}
\end{equation}
where $L$ is the arm length of the detector and $f^*=c/(2\pi L)$ is the transfer frequency.

The inner product between two signals $s_1(t)$ and $s_2(t)$ is defined as
\begin{equation}
\label{gqprod1}
(s_1|s_2)=2\int_0^\infty\frac{\tilde{s_1^*}\tilde{s_2}+\tilde{s_2^*}\tilde{s_1}}{S_n(f)}df,
\end{equation}
where $\tilde{s_1}(f)$ and $\tilde{s_2}(f)$ are the Fourier transforms of the respective gravitational waveforms $s(t)$ registered in the detector;
the noise spectral density $S_n(f)$ for space-based GW detectors is \cite{Robson:2018ifk}
\begin{equation}
S_n(f)=\frac{1}{L^2}\left\{S_x+\left[1+\left(\frac{0.4\text{ mHz}}{f}\right)^2\right]\frac{4S_a}{(2\pi f)^4}\right\}+S_c(f),
\end{equation}
where the confusion noise is \cite{Robson:2018ifk}
\begin{equation}
\begin{split}
S_c(f)=&A f^{-7/3}\exp[-f^\alpha+\beta f \sin(\kappa f)][1\\
&+\tanh (\gamma(f_\kappa-f))],
\end{split}
\end{equation}
and the arm length $L$, the position noise $\sqrt{S_x}$ and the acceleration noise $\sqrt{S_a}$ are the parameters of the detector. $L=2.5\times 10^9$ m, $\sqrt{S_x}=15\ \text{pm/Hz}^{1/2}$ and $\sqrt{S_a}=3\times10^{-15}\ \text{m s}^{-2}/\text{Hz}^{1/2}$ for LISA \cite{Danzmann:1997hm,LISA:2017pwj}, $L=\sqrt{3}\times 10^8$ m, $\sqrt{S_x}=1\  \text{pm/Hz}^{1/2}$ and $\sqrt{S_a}=10^{-15}\ \text{m s}^{-2}/\text{Hz}^{1/2}$ for TianQin \cite{TianQin:2015yph}; and $L=3\times 10^9$ m, $\sqrt{S_x}=8\ \text{pm/Hz}^{1/2}$ and $\sqrt{S_a}=3\times10^{-15}\ \text{m s}^{-2}/\text{Hz}^{1/2}$ for Taiji \cite{Hu:2017mde}.
The SNR for a given GW waveform $h$ is
\begin{equation}
\rho[h]=(s[h]|s[h])^{1/2}.
\end{equation}

We use the FIM method to estimate the source's parameters $\theta^a$.
The parameter error is estimated as
\begin{equation}
\Delta\theta^a=\sqrt{\langle (\theta^a)^2-\langle \theta^a \rangle^2\rangle}\approx \sqrt{\Sigma_{aa}},
\end{equation}
where $\Sigma_{aa}=(\Gamma^{-1})_{aa}$  is the diagonal element of the inverse of the FIM in the large SNR limit,
and the element of FIM $\Gamma_{ab}$ is defined as
\begin{equation}
\Gamma_{ab}\equiv\left(\frac{\partial s[h]}{\partial\theta^a}\left|\frac{\partial s[h]}{\partial\theta^b}\right.\right).
\end{equation}
The covariant matrix between two parameters $\theta^a$ and $\theta^b$ is given by
\begin{equation}
C_{ab}=\Sigma_{ab}/\sqrt{\Sigma_{aa}\,\Sigma_{bb}}.
\end{equation}
The angular resolution $\Delta \Omega_s$ is defined as
\begin{equation}
\Delta \Omega_s = 2\pi\sin\theta_s
\sqrt{\Sigma_{\theta_s\theta_s}\Sigma_{\phi_s\phi_s}-\Sigma^2_{\theta_s\phi_s}}.
\end{equation}
To carry out the parameter estimation, we use the frequency-domain waveform model IMRPHENOMXHM \cite{Pratten:2020fqn,Garcia-Quiros:2020qpx}
\begin{equation}
\tilde{h}(f)=\tilde{h}_+-i\tilde{h}_\times=\sum_{\ell,m}
\tilde{h}_{\ell m}(f)\,_{-2}Y_{\ell m}(\iota,0),
\end{equation}
which describes the mode of nonprecessing coalescing black holes in terms of piecewise closed form expressions,
where $_{-2}Y_{\ell m}(\iota,0)$ are spherical harmonics of spin weight -2,
\begin{equation}
\tilde{h}_{\ell -m}(f)=A_{\ell m}(f)e^{-i\Phi_{\ell m}(f)},
\end{equation}
and the amplitude $A_{\ell m}$ and the phase $\Phi_{\ell m}$ are given by IMRPHENOMXHM waveform model \cite{Garcia-Quiros:2020qpx}.
Note that
\begin{equation}
\begin{split}
\tilde{h}_{\ell m}(f)=(-1)^\ell\tilde{h}^*_{\ell -m}(-f),\\
\tilde{h}_+(f)=\frac{1}{2}[_{-2}Y_{\ell -m}+(-1)^{\ell}\,_{-2}Y^*_{\ell m}]\tilde{h}_{\ell -m}(f),\\
\tilde{h}_\times(f)=\frac{i}{2}[_{-2}Y_{\ell -m}-(-1)^{\ell}\,_{-2}Y^*_{\ell m}]\tilde{h}_{\ell -m}(f).
\end{split}
\end{equation}
For the dependence on the inclination angle $\iota$,
we get $\tilde{h}_+^{(2,2)}(f)\propto (1+\cos^2\iota)/2$ and
$\tilde{h}_\times^{(2,2)}(f)\propto \cos\iota$.
In this paper, we use the $(\ell, |m|)=(2,2),(2,1),(3,3),(4,4)$ modes
to constrain the BD theory and massive gravity.
Because GW interferometers are more sensitive to phase evolution \cite{Chatziioannou:2012rf},
we mainly discuss the phase correction.
The modification of the phase $\Phi_{\ell m}$ in modified gravity is \cite{Chatziioannou:2012rf,Pratten:2020fqn,Garcia-Quiros:2020qpx}
\begin{equation}
\begin{split}
\label{phase}
\tilde{\Phi}_{\ell m}(f)&=\Phi_{\ell m}(f)+\delta\Phi_{\ell m}(f),\\
\delta\Phi_{\ell m}(f)&=\frac{m}{2}\delta\Phi_{22}\left(\frac{2}{m}f\right).
\end{split}
\end{equation}
For Brans-Dicke theory, the amplitude and phase modifications for the (2,2) mode are \cite{Will:1994fb,Yunes:2011aa,Chatziioannou:2012rf,Liu:2020moh}
\begin{equation}
\begin{split}
\label{phase1}
\delta A_{22}=&-\frac{5\mathcal{S}^2 u^{-2/3}}{96(2+\omega_{BD})}=-\frac{7}{8}b u^{-2/3},\\
\delta\Phi_{22}(f)=&-
\frac{3}{128\eta}bu^{-7/3}\left(1-\frac{7}{2}u^{2/3} +5\pi u\right.\\
&\left.-\frac{350}{9}u^{4/3}+\frac{84}{5}\pi u^{5/3}\right),
\end{split}
\end{equation}
where $u=\pi M f$, $f$ is the frequency of the GW;
the total mass $M=M_1+M_2$, $M_1$ and $M_2$ are the masses of the binary;
the difference of the sensitivity between the two bodies in the binaries is assumed to be $\mathcal{S}=0.3$ for NS/BH binaries \cite{Berti:2004bd};
$\omega_{BD}$ is the BD coupling constant; and
\begin{equation}
b=\frac{5\mathcal{S}^2}{84(2+\omega_{BD})}.
\end{equation}
The waveform for the breathing mode in Brans-Dicke theory is
\begin{equation}
\label{bdbth}
\begin{split}
\tilde{h}_b(f)=&\left(\frac{5\pi}{384}\right)^{1/2}\frac{\mathcal{S}}{2+\omega_{BD}}\frac{\mathcal{M}^2}{d_L}\eta^{-7/10}u^{-3/2}\\
&\times\left(\sin\iota e^{-i\tilde{\Phi}_{1,1}(f)}+ 2\sin^2\iota u^{1/3}e^{-i\tilde{\Phi}_{2,2}(f)}\right),
\end{split}
\end{equation}
where the first term in the frequency-domain waveform is  the (1,1) mode \cite{Chatziioannou:2012rf},
the second term is the (2,2) mode,
$d_L$ is the luminosity distance, $\mathcal{M}=\eta^{3/5}M$ is the chirp mass and the symmetric mass ratio $\eta=M_1 M_2/M^2$.
Comparing the waveforms for the tensor and breathing modes,
we see that the amplitude of the breathing mode is suppressed by a factor of $\sim 1/\omega_{BD}$,
so we expect the effect of the breathing mode to not be significant.

Since the dominant effect of a graviton mass is the dephasing due to the frequency-dependent massive graviton velocity \cite{Will:1997bb,Will:2004xi}, we follow Refs. \cite{Will:1997bb,Will:2004xi,Berti:2004bd,Yagi:2009zm}
to add the following term to the phase for the (2,2) mode,
\begin{equation}
\begin{split}
\label{phase2}
\delta\Phi_{22}(f)&=-\frac{\pi^2D M}{\lambda_g^2(1+z)}u^{-1},
\end{split}
\end{equation}
where $\lambda_g$ is the wavelength of massive graviton,
the effective distance $D$ is
\begin{equation}
D=\frac{1+z}{H_0}\int_{0}^{z}\frac{dz'}{(1+z')^2[\Omega_M(1+z')^3+\Omega_\Lambda]^{1/2}},
\end{equation}
$H_0$ is the Hubble constant, $\Omega_M$ is the current matter energy density and $\Omega_\Lambda$
is the current energy density of dark energy which is taken as the cosmological constant for simplicity.
We use the Planck 2018 results $H_0=67.4\ \text{km s}^{-1}\text{Mpc}^{-1}$, $\Omega_M=0.315$ and $\Omega_\Lambda=0.685$ \cite{Planck:2018jri}.
For alternative theories of gravity,
we should consider extra polarizations, e.g., the breathing polarization in BD theory
and the scalar modes in massive gravity.
In this paper, we only consider the scalar mode to see the impact on parameter estimation, and we parametrize the scalar mode in massive gravity $h_B^{\ell m}(f)$ as $A_{B} h_{\ell -m}(f) Y_{\ell m}(\iota,0)$ \cite{Chatziioannou:2012rf,Takeda:2018uai,Wong:2021cmp,Zhang:2021fha}.

\section{Constraints on alternative theories of gravity}
\label{sec3}

In this section, we estimate the errors of the source's parameters:
the total mass $M$, the symmetric mass ratio $\eta$,
the luminosity distance $d_{L}$, the sky location $(\theta_s,\phi_s)$ and the angular resolution $\Delta\Omega_s$,
the inclination angle $\iota$, the polarization angle $\psi_s$ and
the coalescence phase $\phi_{c}$ at the coalescence time $t_{c}$,
so $\bm \theta=\{M$, $\eta$, $d_L$, $\theta_s$, $\phi_s$, $\psi_s$, $\iota$, $t_c$, $\phi_c\}$.
The mass ratio is defined as $q=M_1/M_2$ with $M_1\geq M_2$, so the total mass $M=(1+q)M_2$.
To constrain the graviton mass, we need to add the parameter $\lambda_g$ and $A_B$ if we include the scalar modes.
The effects of massive gravitons are  manifested not only in the phase \eqref{phase} of GW waveforms but also in the transfer function \eqref{transferfunction}.
To constrain BD theory, we need to add the parameter $\omega_{BD}$.
Following Refs. \cite{Berti:2004bd,Yagi:2009zm},
for the bound on graviton mass we consider BBHs with masses between $10^3$ and $10^7$ solar masses at 3 Gpc, and 
for the constraint on the BD coupling constant $\omega_{BD}$,
we use sources of neuron stars inspiraling into intermediate-mass BHs with masses $400\ M_\odot$ and $1000\ M_\odot$.
The massive BBH merger rates could be several hundreds per year \cite{Klein:2015hvg},
so between a few and several tens of massive BBH events per year could be detected with space-based detectors \cite{Klein:2015hvg,LISACosmologyWorkingGroup:2022jok,Wang:2019tto}.
There might be 3-300 massive BBH merger events per year \cite{Sesana:2008ur} or $\leq 0.34$ per year \cite{Li:2022fno}, or $\sim 0.5-1~\rm{yr}^{-1}$ for massive BHs with masses greater than $10^5 M_{\odot}$ \cite{Katz:2019qlu}.
For a four-year mission, depending on the seed model and assumptions, it was estimated that around 1000 massive BBH events could be detected by space-based detectors in the most optimistic scenario \cite{Ricarte:2018mzn}.
The event rate of a NS inspiraling into an intermediate-mass BH is uncertain.
The NS/BH merger rate is estimated to be between 7.8 and 140 Gpc$^{-3}$yr$^{-1}$ using GW events through GWTC-3 \cite{KAGRA:2021duu}.
LISA might detect ten intermediate mass ratio inspirals (IMRIs)
consisting of BHs with $10^3\,M_\odot$ and $10\,M_\odot$ at any given time \cite{Miller:2001ez},
or IMRIs with an event rate $\sim 3-10$ Gpc$^{-3}$ yr$^{-1}$ \cite{Fragione:2017blf}.
To evaluate the capability of space-based detectors on the test of alternative theories of gravity,
for each source with the specific total mass $M$ and redshift $z$, we use Monte Carlo simulation to generate $1000$ sources with $\{\cos\theta_s,\cos\iota\}$ uniformly distributed in $[-1, 1]$ and $\{\phi_s,\psi_s,\phi_c\}$ uniformly distributed in $[0, 2\pi]$.
We calculate the error of each parameter from each binary and then take the average of the 1000 results.
For the observations, we consider one-year observations before the coalescence with the space-based GW detectors LISA, Taiji and TianQin and their combined network.

\subsection{Bound on graviton mass}

\begin{figure*}
$\begin{array}{cc}
\includegraphics[width=0.46\textwidth]{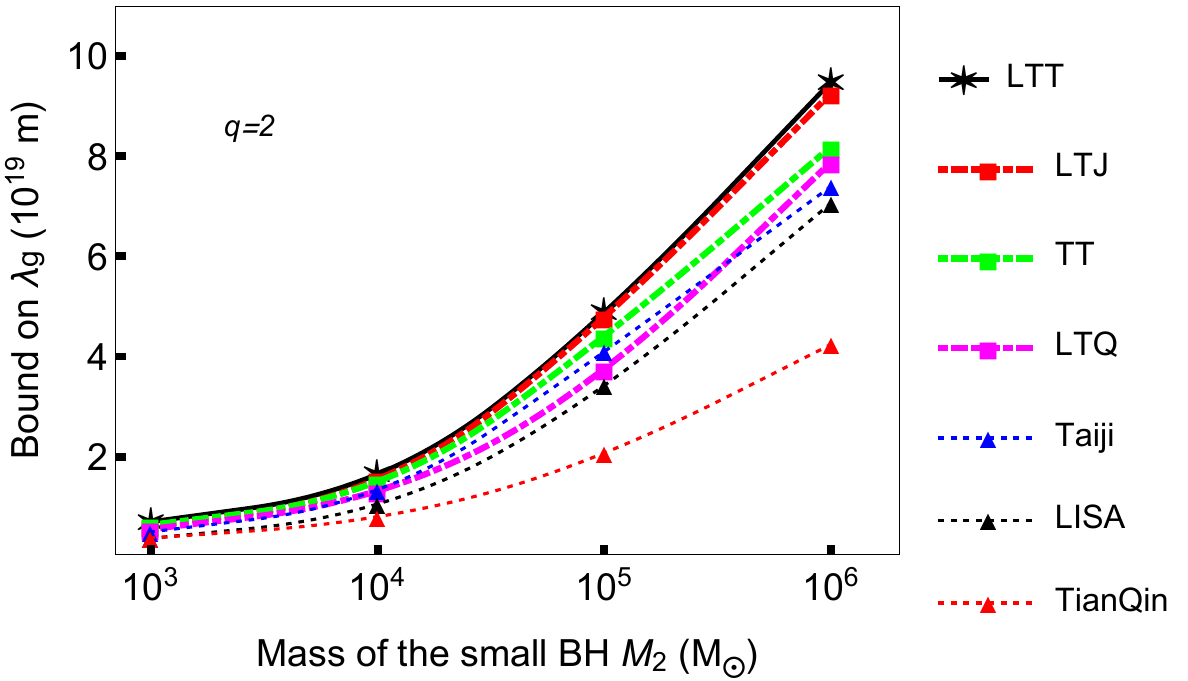}&
\includegraphics[width=0.46\textwidth]{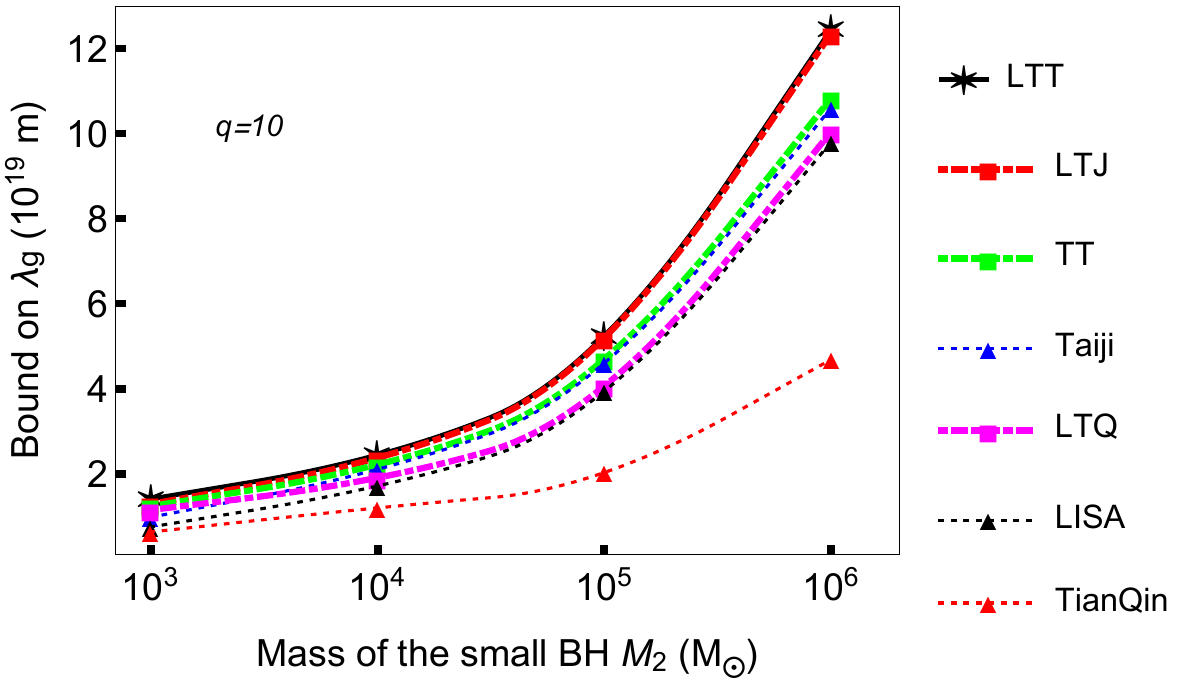}
\end{array}$
\caption{The lower bounds on the wavelength $\lambda_g$ of massive graviton for different GW detectors and their combinations without the scalar mode.
The left panel is for $q=2$ and the right panel is for $q=10$.
The luminosity distance $d_L$ is chosen as 3 Gpc.
Different lines denote different detectors and different detector network.}
\label{fig1}
\end{figure*}

\begin{figure*}
$\begin{array}{cc}
\includegraphics[width=0.46\textwidth]{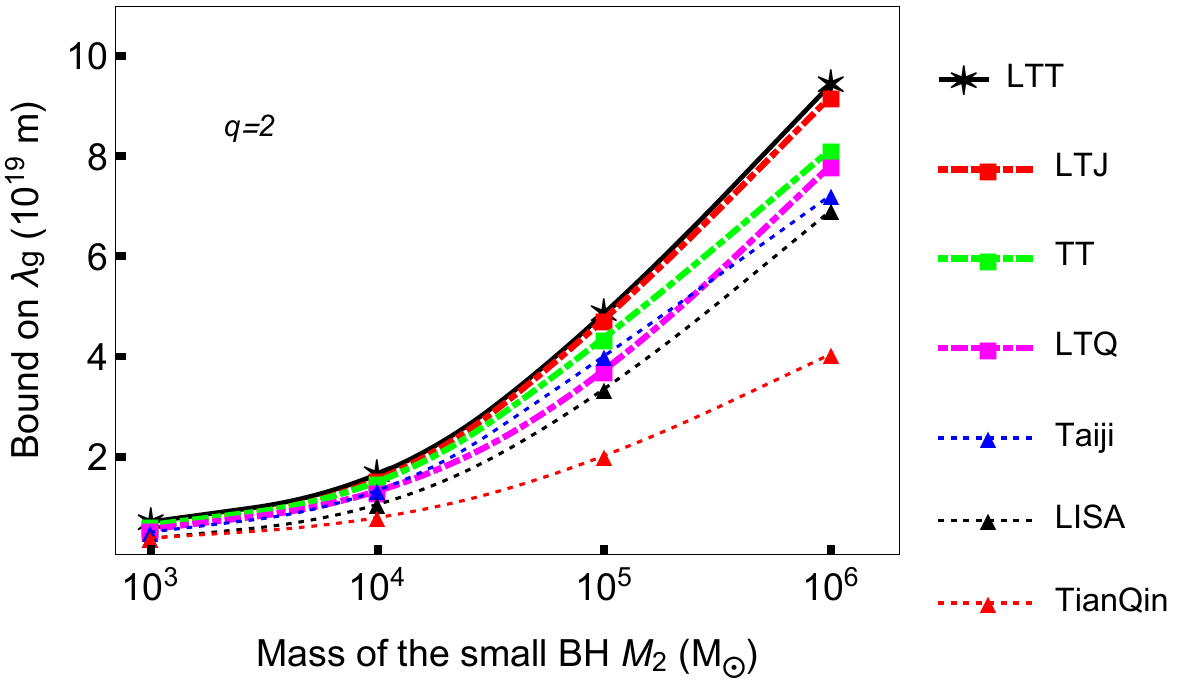}&
\includegraphics[width=0.46\textwidth]{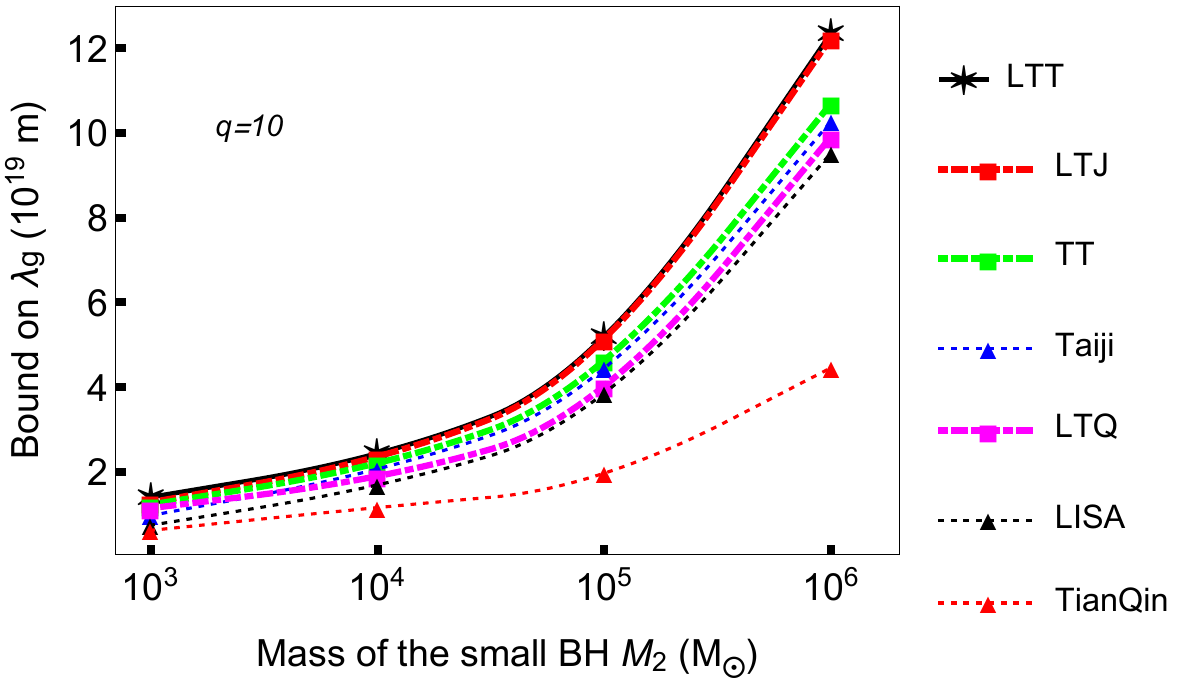}
\end{array}$
\caption{The lower bounds on the wavelength $\lambda_g$ of massive graviton for different GW detectors and their combinations in massive gravity with the scalar mode.
The left panel is for $q=2$ and the right panel is for $q=10$.
The luminosity distance $d_L$ is 3 Gpc. Different lines denote different detectors.}
\label{fig2}
\end{figure*}

\begin{figure*}
$\begin{array}{cc}
\includegraphics[width=0.46\textwidth]{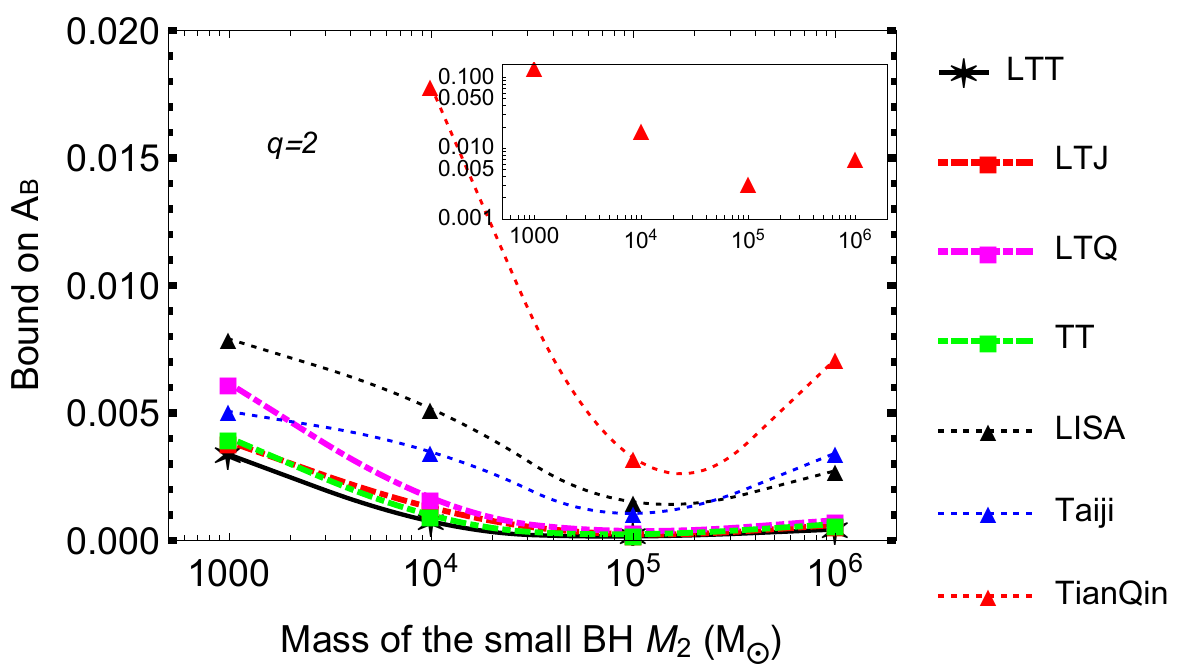}&
\includegraphics[width=0.46\textwidth]{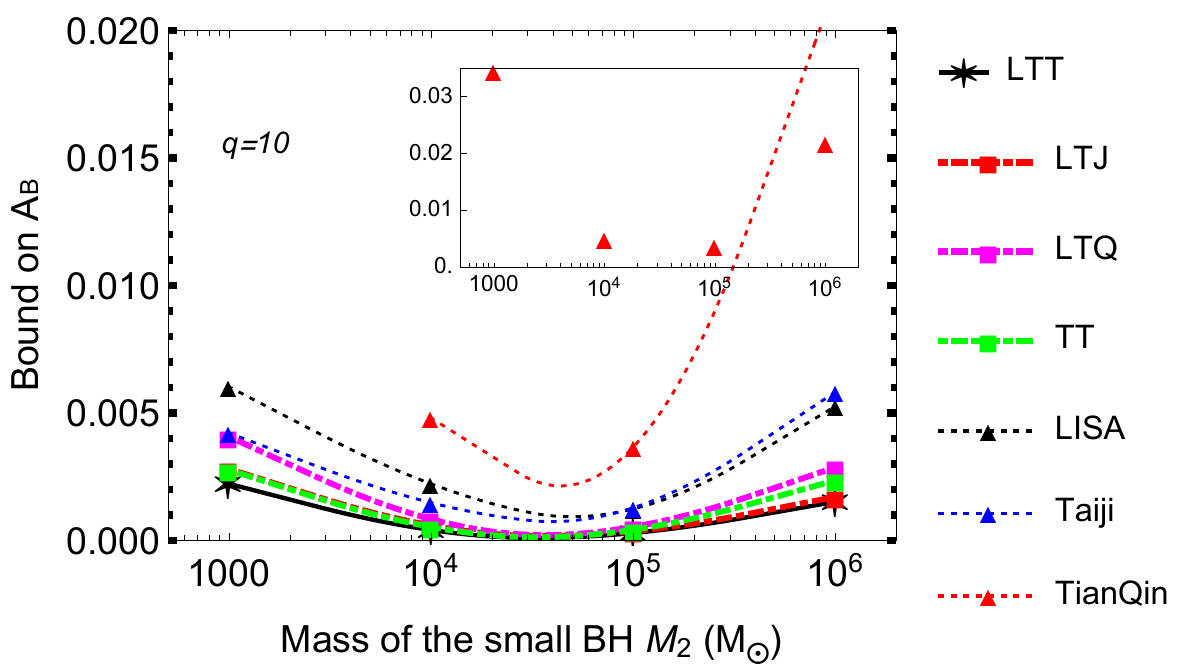}
\end{array}$
\caption{The lower bounds on the relative amplitude $A_B$ of the scalar mode for different GW detectors and their combinations in massive gravity with the scalar mode.
The left panel is for $q=2$ and the right panel is for $q=10$.
The luminosity distance $d_L$ is 3 Gpc.
Different lines denote different detectors. The insets show the results for TianQin.}\label{fig3}
\end{figure*}

For the constraint on the mass of graviton,
we consider BBHs with different masses at 3 Gpc.
We choose the masses of the small BHs as $10^6\ M_{\odot}$, $10^5\ M_{\odot}$, $10^4\ M_{\odot}$, and $10^3\ M_{\odot}$
and two different mass ratios, $q=2$ and $q=10$.
To evaluate the capability of GW detector on the constraint of the wavelength $\lambda_g$ of massive gravitons,
we consider all three space-based GW detectors LISA, Taiji, and TianQin and their combinations. For the sake of convenience,
we denote the network of LISA and Taiji as LTJ, of LISA and TianQin as LTQ,
of TianQin and Taiji as TT,
and of LISA, Taiji and TianQin as LTT.
We also consider the impact of the scalar mode on the constraint of the wavelength $\lambda_g$ by including the relative amplitude $A_B$ of the scalar mode.
The errors of the parameter estimation and the constraints on $\lambda_g$ and $A_B$
are shown in Tables \ref{table1} and \ref{table2} and Figs. \ref{fig1}, \ref{fig2}, and \ref{fig3}.

In Table \ref{table1}, we show the estimation errors of all parameters along with SNRs for BBHs with masses $(10^5+10^6)M_\odot$ at 3 Gpc with different GW detectors and their combinations.
From Table \ref{table1}, we see that while the error of angular resolution $\Delta\Omega_s$ with TianQin is almost 3 orders of magnitude worse than those with LISA or Taiji,
the constraint on $\lambda_g$ is only $\sim 2$ times worse and the constraint on $A_B$ is only $\sim 3$ times worse.
The inclusion of the scalar mode has little effect on the constraint on $\lambda_g$.
Although the networks LTJ, LTQ, TT, and LTT greatly improve the accuracy of sky localization,
the improvements on the bound on $\lambda_g$ are small compared with LISA or Taiji alone.
The results for BBHs with other masses are similar.
In Table \ref{table2}, we show the constraints on $\lambda_g$ and $A_B$ along with the values of SNR for different BBHs with the network LTT. The results with other detectors are similar.
The results show that the inclusion of the scalar mode has little effect on the constraint on $\lambda_g$.
As the total mass of the BBH increases, the bound on $\lambda_g$ becomes stronger.
The constraint on $A_B$ and the other parameters become better as the total mass of BBH increases to a certain value ,
but the bound on $A_B$ and the constraints on the other parameters along with the SNR then decrease as the total mass further increases.

In Figs. \ref{fig1} and \ref{fig2} we show the bounds on $\lambda_g$ versus the mass of the small BH of the BBH for different detectors with and without the scalar mode, respectively.
In Fig. \ref{fig3}, we show the bounds on $A_B$ versus the mass of the small BH of the BBH with $q=2$ and $q=10$ for different detectors.
From Figs. \ref{fig1} and \ref{fig2}, we see that the addition of the scalar mode has little effect on the bound on $\lambda_g$ for all BHHs and all detectors.
As the total mass of BBHs increases, the bound on $\lambda_g$ becomes stronger for all detectors.
Although the constraints on other parameters are much worse with TianQin,
the bound on $\lambda_g$ with TianQin is just at most $\sim 2$ times weaker than LISA or Taiji.
The improvements on the bounds on $\lambda_g$ and $A_B$ with the networks LTJ, LTQ, TT and LTT are small compared with LISA, Taiji, and TianQin alone.
The strongest bound is obtained for BBHs with masses $(10^6+10^7)M_\odot$ with LTT, and the result is $\lambda_g\gtrsim 1.24 \times 10^{20}$ m.
For TianQin, BBHs with the total mass around $3\times 10^5M_\odot$
give the largest SNR and the best constraints on $A_B$ and the other parameters.
For LISA, Taiji and the networks LTQ, LTJ, and LTT,
BBHs with the total mass around $5\times 10^5M_\odot$ give the largest SNR and the best constraints on $A_B$ and the other parameters, the strongest bound for $A_B$ is $A_B< 5.7\times 10^{-4}$.
The dependence of the constraints on the total mass is similar for different  mass ratio $q$.

For the one-year observations of BBHs with masses $(1+1)\times 10^6M_\odot$ at 3 Gpc with LISA, it was found that $\lambda_g>4.95\times 10^{19}$ m
and $\lambda_g>1.33\times 10^{19}$ m if the spin-orbit coupling is included
\cite{Berti:2004bd}.
For the one-year observations of
BBHs with masses $(1+10)\times 10^6M_\odot$ at 3 Gpc with LISA, including both the spin-orbit and spin-spin couplings, it was found that $\lambda_g>4.1\times 10^{18}$ m and
$\lambda_g>4.9\times 10^{19}$ m with the spin precession \cite{Yagi:2009zm}.
Using the pattern-averaged analysis for BBHs with masses $(1+1)\times 10^6M_\odot$ at 3 Gpc,
it was obtained that $\lambda_g>1.82\times 10^{19}$ m with LISA,
$\lambda_g>1.91\times 10^{19}$ m with Taiji,
and $\lambda_g>1.1\times 10^{19}$ m with TianQin \cite{Gao:2022hho}.
In this paper, because we take higher modes into account, we get $\lambda_g>7.07\times 10^{19}$ m for BBHs with masses $(1+2)\times 10^6M_\odot$ at 3 Gpc with LISA,
and $\lambda_g>9.80\times 10^{19}$ m for BBHs with masses $(1+10)\times 10^6M_\odot$ at 3 Gpc with LISA.
For the one-year observations of BBHs with masses $(1+10)\times 10^6M_\odot$ at 3 Gpc with LTT, we get $\lambda_g\gtrsim 1.24 \times 10^{20}$ m.

\begin{table*}[htbp]
	\renewcommand\tabcolsep{4.0pt}
    \centering
	\begin{tabular}{llllllllll}
		\hline
		\hline
		Detectors & SNR  &$\Delta\ln M$ & $\Delta \ln \eta$ & $\Delta \ln d_L$ & $\Delta\Omega_s$ & $\Delta\psi_s$ & $\Delta \iota$ &  $\lambda_g$& $A_B$\\
		& &  ($10^{-4}$)& ($10^{-5}$) &($10^{-2}$) &($10^{-5}$str)& ($10^{-2}$) &($10^{-3}$)  &($10^{19}\text{m}$)&($10^{-3}$)\\

		\hline
 		LISA  &$1583$ ~& $1.78$~& $2.19$~&$1.00$~&$57.4$~&$2.01$~&$1.93$~&$3.94$~&$\cdots$\\
 		~  &$1583$ ~& $1.85$~& $2.28$~&$1.04$~&$66.6$~&$2.05$~&$2.64$~&$3.87$~&$4.43$\\
        Taiji  &$1992$ ~& $1.33$~& $1.63$~&$0.75$~&$41.3$~&$1.64$~&$1.42$~&$4.60$~&$\cdots$\\
        ~  &$1992$ ~& $1.42$~& $1.74$~&$0.78$~&$49.0$~&$1.68$~&$1.95$~&$4.46$~&$4.58$\\
        TianQin  &$490$ ~& $6.88$~& $8.40$~&$6.86$~&$10400$~&$11.0$~&$6.54$~&$2.03$~&$\cdots$\\
        ~  &$490$ ~& $7.33$~& $8.94$~&$7.05$~&$12490$~&$11.1$~&$9.90$~&$1.97$~&$13.2$\\
        LTJ  &$2574$ ~& $1.02$~& $1.25$~&$0.10$~&$0.14$~&$0.16$~&$0.76$~&$5.20$~&$\cdots$\\
        ~  &$2574$ ~& $1.04$~& $1.27$~&$0.11$~&$0.15$~&$0.16$~&$0.78$~&$5.15$~&$1.29$\\
        LTQ  &$1674$ ~& $1.66$~& $2.04$~&$0.18$~&$0.97$~&$0.27$~&$1.26$~&$4.06$~&$\cdots$\\
        ~  &$1674$ ~& $1.69$~& $2.07$~&$0.18$~&$1.02$~&$0.28$~&$1.30$~&$4.03$~&$1.91$\\
        TT  &$2073$ ~& $1.26$~& $1.54$~&$0.14$~&$0.68$~&$0.25$~&$1.05$~&$4.69$~&$\cdots$\\
        ~  &$2073$ ~& $1.29$~& $1.58$~&$0.14$~&$0.71$~&$0.27$~&$1.11$~&$4.64$~&$1.51$\\
        LTT  &$2633$ ~& $1.00$~& $1.22$~&$0.09$~&$0.10$~&$0.14$~&$0.72$~&$5.24$~&$\cdots$\\
        ~  &$2633$ ~& $1.01$~& $1.24$~&$0.09$~&$0.10$~&$0.14$~&$0.73$~&$5.21$~&$1.06$\\

		\hline
		\hline
	\end{tabular}
	\caption{The results of error estimation for different GW detectors and their combinations in massive gravity with and without the scalar mode.
 The masses of the BBHs are $10^6M_{\odot}$ and $10^5M_{\odot}$, and the luminosity distance $d_L$ is 3 Gpc.
The first line of each detector shows the  error estimation without considering the scalar mode.
LTJ denotes the LISA+Taiji network, LTQ denotes the LISA+TianQin network, TT denotes the TianQin+Taiji network, and LTT denotes the LISA+Taiji+TianQin network.}
\label{table1}
\end{table*}

\begin{table*}[htbp]
	\renewcommand\tabcolsep{4.0pt}
    \centering
	\begin{tabular}{llll|llll}
		\hline
		\hline
		\multicolumn{4}{c|}{$q=M_1/M_2=2$} & \multicolumn{4}{c}{$q=M_1/M_2=10$}\\
		\hline
			$M_2$  & SNR    & $\lambda_g$& $A_B$ & $M_2$ & SNR   & $\lambda_g$& $A_B$\\
			($M_{\odot}$)& &($10^{19}\text{ m}$)&($10^{-3}$)&($M_{\odot}$)& &($10^{19}\text{ m}$)&($10^{-3}$)\\\hline
 		$10^6$   &$1892$ ~&$9.49$~&$\cdots$&$10^6$ &$524$ ~& $12.4$~&$\cdots$\\
 		~   &$1892$ ~&$9.44$~&$1.52$&~ &$524$ ~& $12.4$~&$5.44$\\
        $10^5$   &$6002$~&$4.89$~&$\cdots$&$10^5$ &$2633$~&$5.24$~&$\cdots$\\
        ~   &$6002$~&$4.87$~&$0.57$&~ &$2633$~&$5.21$~&$1.06$\\
        $10^4$   &$1156$ ~&$1.66$~&$\cdots$&$10^4$ &$1910$ ~&$2.44$~&$\cdots$\\
        ~   &$1156$~&$1.65$~&$2.59$&~ &$1910$~&$2.44$~&$1.44$\\
        $10^3$   &$170$ ~&$0.70$~&$\cdots$&$10^3$ &$309$ ~&$1.41$~&$\cdots$\\
        ~   &$170$~&$0.70$~&$11.7$&~ &$309$~&$1.40$~&$7.71$\\

		\hline
		\hline
	\end{tabular}
	\caption{The results of the lower bounds on the wavelength $\lambda_g$ of massive graviton and the relative amplitude $A_B$ of the scalar mode with the LISA+TianQin+Taiji network (LTT). We consider BBHs with masses $M_2$ for the small BHs $10^6M_{\odot}$, $10^5M_{\odot}$, $10^4M_{\odot}$ and $10^3M_{\odot}$, and the mass ratio $q=2$ and $q=10$ at $3$ Gpc. We also show the results of SNR for each binary. The first line of each binary shows the results without considering the scalar mode.}
\label{table2}
\end{table*}

\subsection{Constraint on BD coupling constant}
BHs do not carry scalar charges due to the no-hair theorem,
and the sensitivity of BHs is always $s_\text{BH}=0.5$,
so we cannot use BBHs to constrain BD theory via dipole radiation.
For this reason, we consider NS inspiraling into intermediate-mass BHs.
It was found that NS inspiraling into the intermediate-mass BH with smaller mass gives better constraint on $\omega_{BD}$
because more GW cycles are observed for the same observational time \cite{Scharre:2001hn,Will:2004xi}.
Following Refs. \cite{Scharre:2001hn,Will:2004xi,Berti:2004bd,Yagi:2009zm},
we fix the mass of the NS as $M_{2}=1.4M_\odot$ and assume its sensitivity $s_\text{NS}=0.2$
and consider intermediate-mass BHs with the mass $400\ M_{\odot}$ and $1000\ M_{\odot}$,
so $\mathcal{S}=s_\text{BH}-s_\text{NS}=0.3$.
For each space-based GW detector, we use both Michelson interferometers I and II and consider binaries of $\rho=\sqrt{200}$;
the luminosity distance and redshift are then determined based on the flat $\Lambda$CDM model
with the cosmological parameters $H_0=67.4$ km/s/Mpc, $\Omega_M=0.315$, and $\Omega_\Lambda=1-\Omega_M=0.685$ \cite{Planck:2018jri}.
For the network of GW detectors including LISA, we set the SNR of LISA as $\rho=\sqrt{200}$ and the total SNR of the network is the sum of that from each individual detector.
For the network TT, we set the SNR of Taiji as $\rho=\sqrt{200}$, and the total SNR of the network TT is the sum of SNRs from Taiji and TianQin.
The results of the average error of the parameter estimation in BD theory for $1.4/1000\ M_{\odot}$ NS/BH binaries with different detectors and their combinations  are shown in Table \ref{bd}.
The average lower bounds on $\omega_{BD}$ with the network LTT are shown in Table \ref{bd1}.

In Table \ref{bd}, we show the estimation errors of all parameters along with SNRs for NS/BH binaries with masses $1.4/1000\ M_{\odot}$ by different GW detectors and their combinations.
The results for $1.4/400\ M_{\odot}$ NS/BH binaries are similar.
From Table \ref{bd}, we see that the breathing mode has negligible effect on the constraint on $\omega_{BD}$.
The constraints with Taiji are a little better than those with LISA.
For some parameters, Taiji gives better constraints than TianQin;
but for other parameters, TianQin gives better constraints.
Compared with single detector, the network improves the accuracy of source parameters including the sky localization by several times to even 1 order of magnitude than single detector,
and the improvement on the bound on $\omega_{BD}$ is about $\sim 2-4$ times.
In Table \ref{bd1}, we show the constraints on $\omega_{BD}$ and the values of SNR for the two NS/BH binaries with the network LTT.
The results also show that the inclusion of the breathing mode has negligible effect on the constraint on $\omega_{BD}$ because the breathing mode is suppressed by $1/\omega_{BD}$.
As expected, the bound on $\omega_{BD}$ and the constraints on source parameters from $1.4/400\ M_{\odot}$ NS/BH binaries are a little stronger than that from $1.4/1000\ M_{\odot}$ NS/BH binaries because the SNR is larger.
With and without the breathing mode, the bounds on $\omega_{BD}$ with LTT are the same and the result is $\omega_{BD}>6.11\times10^{6}$ for $1.4/400\ M_{\odot}$ NS/BH binaries.

For the one-year observations of $1.4/1000\ M_{\odot}$
NS/BH binaries with LISA, it was found
that $\omega_{BD}>96719$ \cite{Berti:2004bd}.
If the spin-orbit and spin-spin couplings were included,
the constraint became $\omega_{BD}>4844$
and $\omega_{BD}>3523$ for precessing eccentric binaries \cite{Yagi:2009zm}.
Our result $\omega_{BD}>5.17\times10^{6}$ and the constraints on source parameters given in Table \ref{bd} are better because we consider higher modes for the waveform.

The GW frequencies from the merger of $1.4/400\ M_{\odot}$ and $1.4/1000\ M_{\odot}$ NS/BH binaries are in the frequency band of ground-based detectors,
so we can improve the bound $\omega_{BD}$ by combining the measurements from space-based detectors and ground-based detectors; i.e.,
we expect that the bound can be improved with a multiband analysis.

\begin{table*}[htbp]
	\renewcommand\tabcolsep{4.0pt}
    \centering
	\begin{tabular}{lllllllll}
		\hline
		\hline
		Detectors & SNR  &$\Delta\ln M$ & $\Delta \ln \eta$ & $\Delta \ln d_L$ & $\Delta\Omega_s$ & $\Delta\psi_s$ & $\Delta \iota$  & $\omega_{BD}$\\
		& &  ($10^{-7}$)& ($10^{-8}$) &  &($10^{-5}$ str)&  &  &($10^{6}$)  \\
		\hline
 		 LISA  &$14$ ~& $2.70$~& $7.75$~&$0.19$~&$4.72$~&$0.17$~&$0.15$~&$1.39$~\\
 		 ~  &$14$ ~& $2.70$~& $7.75$~&$0.19$~&$4.72$~&$0.17$~&$0.15$~&$1.39$~\\
        Taiji  &$14$ ~& $2.35$~& $7.57$~&$0.19$~&$3.65$~&$0.17$~&$0.15$~&$1.38$~\\
        ~  &$14$ ~& $2.35$~& $7.57$~&$0.19$~&$3.66$~&$0.17$~&$0.15$~&$1.38$~\\
        TianQin  &$14$ ~& $2.39$~& $7.35$~&$0.28$~&$5.92$~&$0.33$~&$0.22$~&$1.40$~\\
        ~  &$14$ ~& $2.40$~& $7.36$~&$0.28$~&$5.92$~&$0.33$~&$0.22$~&$1.40$~\\
        LTJ  &$30$ ~& $0.83$~& $3.57$~&$0.09$~&$0.61$~&$0.08$~&$0.07$~&$2.96$~\\
        ~  &$30$ ~& $0.83$~& $3.58$~&$0.09$~&$0.60$~&$0.08$~&$0.07$~&$2.96$~\\
        LTQ  &$44$ ~& $0.79$~& $2.59$~&$0.07$~&$0.49$~&$0.07$~&$0.06$~&$4.39$~\\
        ~  &$44$ ~& $0.79$~& $2.59$~&$0.07$~&$0.49$~&$0.07$~&$0.06$~&$4.39$~\\
        TT  &$27$ ~& $1.14$~& $4.17$~&$0.11$~&$1.05$~&$0.10$~&$0.08$~&$2.63$~\\
        ~  &$27$ ~& $1.14$~& $4.17$~&$0.11$~&$1.05$~&$0.10$~&$0.08$~&$2.63$~\\
        LTT  &$52$ ~& $0.55$~& $2.13$~&$0.05$~&$0.26$~&$0.05$~&$0.04$~&$5.17$~\\
        ~  &$52$ ~& $0.55$~& $2.13$~&$0.05$~&$0.26$~&$0.05$~&$0.04$~&$5.17$~\\

		\hline
		\hline
	\end{tabular}
	\caption{The results of error estimation for $1.4/1000\ M_{\odot}$ NS/BH binaries by different GW detectors and their combinations in BD theory with and without the breathing mode. We use both Michelson interferometers I and II and consider binaries of $\rho=\sqrt{200}$.
For the network of GW detectors including LISA, we set the SNR of LISA as $\rho=\sqrt{200}$, and the total SNR of the network is the sum of that from each individual detector.
For the network TT, we set the SNR of Taiji as $\rho=\sqrt{200}$, and the total SNR of the network TT is the sum of SNRs from Taiji and TianQin.
	The first line of each detector shows the error estimation without considering the scalar mode, and the second line of each detector shows the error estimation including the scalar mode.}
\label{bd}
\end{table*}

\begin{table*}[htbp]
	\renewcommand\tabcolsep{4.0pt}
    \centering
	\begin{tabular}{l|ll|ll}
		\hline
		\hline
		& \multicolumn{2}{c|}{$1.4/1000\ M_{\odot}$} & \multicolumn{2}{c}{$1.4/400\ M_{\odot}$}\\
		\hline
		&	SNR    & $\omega_{BD}$ &  SNR   & $\omega_{BD}$\\
		&	 &($10^{6}$)&  &($10^{6}$)\\\hline
 		No breathing& $52$ ~&$5.17$~&$62$ ~& $6.11$~\\
 		With breathing& $52$ ~&$5.17$~&$62$ ~& $6.11$~\\

		\hline
		\hline
	\end{tabular}
	\caption{The results of the lower bounds on the $\omega_{BD}$ in BD theory with the LISA+TianQin+Taiji network (LTT). We consider NS/BH binaries with masses $1.4/1000\ M_{\odot}$ and $1.4/400\ M_{\odot}$. We also show the results of SNR for each binary.}
\label{bd1}
\end{table*}

\section{Conclusion}
\label{sec4}

By Monte Carlo simulation of 1000 uniformly distributed binaries across the sky,
we evaluate how accurately the future space-based GW detectors such as LISA, Taiji and TianQin and their combined networks can determine source parameters and constrain alternative theories of gravity.
In the constraint of alternative theories of gravity, we also consider the impact of extra polarization by parametrizing the waveform of the scalar mode $h_B^{\ell m}(f)$ in massive gravity as $A_{B} h_{\ell -m}(f) Y_{\ell m}(\iota,0)$, and including the (1,1) and (2,2) modes for the breathing state in BD theory,
we find that the inclusion of the scalar mode has little effect on the constraints on source parameters, the graviton mass, and the BD coupling constant $\omega_{BD}$.

For the constraint on the graviton mass,
we consider BBHs with different masses at 3 Gpc.
The masses of the small BHs are chosen as $10^6M_{\odot}$, $10^5M_{\odot}$, $10^4M_{\odot}$, and $10^3M_{\odot}$,
and the mass ratio between the two constituent BHs in the binary is chosen as $q=2$ and $q=10$.
The mass of graviton affects not only GW phase but also the transfer function, and we take both effects into account.
Although the error of angular resolution $\Delta\Omega_s$ with TianQin is almost 1 order of magnitude worse than those with LISA or Taiji,
the constraint on $\lambda_g$ is only $\sim 2$ times worse, and the constraint on $A_B$ is only $\sim 3$ times worse.
Although the networks LTJ, LTQ, TT, and LTT greatly improve the accuracy of sky localization,
the improvements on the bound on $\lambda_g$ are small compared with single detector.
As the total mass of the BBH increases, the bound on $\lambda_g$ becomes stronger for all detectors.
The constraint on $A_B$ and the other parameters becomes better as the total mass of BBH increases to a certain value,
but the bound on $A_B$ and the constraints on the other parameters along with the SNR then decrease as the total mass further increases.
The strongest bound is obtained for BBHs with masses $(10^6+10^7)M_\odot$ with LTT, and the result is $\lambda_g\gtrsim 1.24 \times 10^{20}$ m.
For TianQin, BBHs with the total mass around $3\times 10^5M_\odot$
give the largest SNR and the best constraints on $A_B$ and the other parameters.
For LISA, Taiji and the networks LTQ, LTJ and LTT,
BBHs with the total mass around $5\times 10^5M_\odot$ give the largest SNR and the best constraints on $A_B$ and the other parameters, the strongest bound for $A_B$ is $A_B< 5.7\times 10^{-4}$.
The dependence of the constraints on the total mass is similar for different mass ratio $q$.

For the constraints on Brans-Dicke theory,
we consider $1.4M_\odot$ NS inspiralling into intermediate-mass BHs with masses $400\ M_\odot$ and $1000\ M_\odot$.
The constraints with Taiji are a little better than those with LISA.
For some parameters, Taiji gives better constraints than TianQin;
but for other parameters, TianQin gives better constraints.
Comparing with single detector, the network improves the accuracy of source parameters including the sky localization by several times to even one order of magnitude than single detector,
and the improvement on the  bound on $\omega_{BD}$ is about $\sim 2-4$ times.
The bounds on $\omega_{BD}$ and the constraints on source parameters from $1.4/400\ M_{\odot}$ NS/BH binaries are a little stronger than that from $1.4/1000\ M_{\odot}$ NS/BH binaries.
The lower bound on $\omega_{BD}$ is $\omega_{BD}>6.11\times10^{6}$ with LTT  for $1.4/400\ M_{\odot}$ NS/BH binaries.

In conclusion, the inclusion of the scalar mode has little effect on the constraints on source
parameters, the graviton mass, and the BD coupling constant $\omega_{BD}$.
By taking higher modes into account, we can get better constraints on alternative theory of gravity.
The detector network can greatly improve the estimation errors of source parameters,
but the improvement of the constraint on alternative theory of gravity is small.

\begin{acknowledgments}
This work is supported in part by the National Key Research and Development Program of China under Grant No. 2020YFC2201504,
the National Natural Science Foundation of China (Grant No. 12175184),
the Chongqing Natural Science Foundation
(Grant No. CSTB2022NSCQ-MSX1324), and the China Postdoctoral Science Foundation (Grant No. BX20220313).
\end{acknowledgments}



%

\end{document}